\documentclass[superscriptaddress,twocolumn,prl,showpacs,floatfix]{revtex4}
\usepackage{graphicx,graphics,color,epsfig} 
\usepackage{bm}
\usepackage{amssymb}
\usepackage{amsmath}
\usepackage{amsfonts}
\usepackage{subfigure}
\usepackage{units}

\newcommand{\s}{\ensuremath{\mathrm{sgn}}}
\newcommand{\etal}{\textit{et al.}}

\begin{document}

\title{Signature of valley polarization in fractional flux
periodicity of a graphene ring}
\author{D. S. L. Abergel}
\email{abergel@cc.umanitoba.ca}
\affiliation{Department of Physics and Astronomy, University 
of Manitoba, Winnipeg, R3T 2N2, Canada.}
\author{Vadim M. Apalkov}
\affiliation{Department of Physics and Astronomy, Georgia State
University, Atlanta, Georgia 30303, USA.}
\author{Tapash Chakraborty}
\email{chakrabt@cc.umanitoba.ca}
\affiliation{Department of Physics and Astronomy, University 
of Manitoba, Winnipeg, R3T 2N2, Canada.}

\begin{abstract}
We have studied the interplay of valley polarization and the Coulomb
interaction on the energy spectrum, persistent current, and optical
absorption of a graphene quantum ring.
We show that the interaction has a dramatic effect on the nature of the
ground state as a function of the magnetic flux, and that the absence of
the exchange interaction for electrons in opposite valleys means that the
singlet-triplet degeneracy is not lifted for certain states. 
The additional level crossings (fractional flux periodicity) due to
the interaction directly leads to extra steps in the persistent current
and intricate structures in the absorption spectrum that should be
experimentally observable. 
By varying the width of the ring, the nature of the ground state at zero
field can be varied and this is manifested in the measurable properties
we discuss.
\end{abstract}

\maketitle

Graphene, a single layer of carbon atoms arranged in a honeycomb
lattice, with its unusual electronic properties, has claimed the 
center stage of condensed matter research for the past three years
\cite{bib:Graphene-ref}. Theoretically investigated sixty years ago
\cite{bib:Wallace-pr94}, interest in this system widened after
free-standing graphene flakes were obtained experimentally by Geim
\etal~in 2004. 
The linear band structure predicted for this material has been verified
and several striking experimental observations have been made, including
the `half-integer' quantum Hall effect and a solid-state manifestation
of the Klein paradox for massless Dirac fermions. \cite{bib:Geim-NM6}

In bulk graphene, the Fermi energy is located at the two inequivalent
$K$ points and the corresponding pairs of single-particle eigenstates in
each of these `valleys' are degenerate with each other. An experimental
observation \cite{bib:Zhang-prl96} of lifting of this valley degeneracy
in high-field quantum Hall effect measurements has prompted several
theoretical studies seeking the origin of this valley polarization
\cite{bib:tcpp}. The intriguing possibility of controlling the
energy difference between electron states in opposite valleys which
would facilitate the idea of valley-tronics (utilizing the valley
quantum number to control the system) would be an exciting development.

Quantum rings of nanoscale dimensions are known to carry a persistent
current: an equilibrium current driven by the magnetic field threading
the ring. This is a direct consequence of the Aharonov-Bohm (AB)
effect which manifests itself as periodic oscillations in the energy
spectrum of the electronic system as a function of the number of
flux quanta entering the ring \cite{bib:Chakraborty-etal}. Impressive 
progress in fabricating nanosize quantum rings containing only
a few electrons has led to equally notable results on the
observation of the properties of energy spectra, first noticed 
in Ref.~\cite{bib:Chakraborty-etal}, via magneto-absorption
spectroscopy \cite{bib:Lorke_etal} and in magneto-transport measurements
\cite{bib:Fuhrer-Nat413}. The important role of the
electron-electron interaction in this system was found to lift the
degeneracy between states with different spin as a means to gain
the exchange energy \cite{bib:Niemela-epl36}. A direct consequence
of this is the fractional flux periodicity that was indeed
observed in subsequent experiments on semiconductor quantum rings
containing only about four electrons \cite{bib:Keyser-prl90}.

Graphene rings have recently been fabricated and AB oscillations were 
observed in their conductance \cite{bib:Russo-prb77}. 
The combined effect of the ring confinement and applied magnetic flux 
is suggested theoretically to lift the orbital degeneracy arising 
from the two valleys in a controllable way \cite{bib:Recher-prb76}. 
Further, a ring with quantum point contacts has been shown to polarize 
the transport current with respect to the valley \cite{bib:Rycerz-arxiv}. 
The important effects of the Coulomb interaction on the valley degeneracy 
and ground state properties have not yet been investigated however.

In this work, we report on the effect of the Coulomb 
interaction on the energy spectrum, persistent current and optical
absorption spectrum of a graphene quantum ring. We show that the 
interaction, the total valley quantum number and spin will dramatically
change the nature of the ground state of a few-electron system. We find 
that the interaction causes drastic changes in the nature of the ground 
state as the flux varies and that the absence of the exchange interaction 
for electrons in opposite valleys means that the singlet-triplet
degeneracy is not lifted for some states. The extra crossings in the
spectrum which are generated by the interaction manifest themselves as
steps appearing in the persistent current (the fractional AB effect) and
results in intricate structures in the absorption spectrum. These
effects are all experimentally measurable. Details of the periods of the
oscillations of the persistent current depend on the width of the ring
and hence the interplay between kinetic and Coulomb energies.

\begin{figure}[tb]
	\centering
	\includegraphics{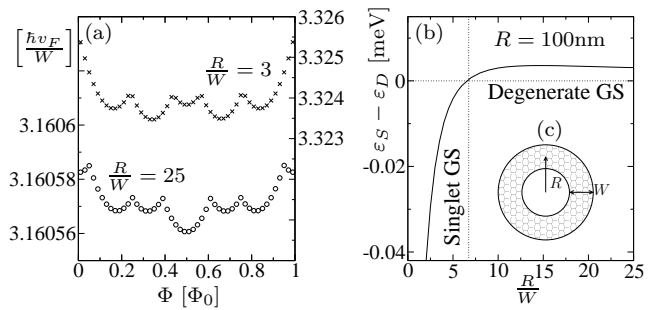}
	\caption{The effect of the ring width on the ground state energy.
	(a) The $\frac{R}{W}=3$ curve (crosses) is plotted relative to the
	right-hand axis, and the $\frac{R}{W}=25$ curve (circles) relative to
	the left-hand axis. 
	(b) The energy difference between lowest singlet ($\varepsilon^{}_S$) and
	degenerate singlet-triplet ($\varepsilon^{}_D$) states at
	$\Phi/\Phi^{}_0=0$. 
	(c) The geometry of the ring.
	\label{fig:Wcomp}}
\end{figure}

We use the valley-symmetric from of the graphene Hamiltonian 
\cite{bib:Recher-prb76}
\begin{equation}
	\label{eq:Hamiltonian}
	{\cal H} = \tau^{}_0 \otimes {\cal H}^{}_0 + \tau^{}_z \otimes
	\sigma^{}_z V(r) 
\end{equation}
where $V(r)$ is a mass term which describes the confinement of the
electron, ${\cal H}^{}_0 = v^{}_F\left(\vec{p}\cdot\vec{\sigma}\right)$
is the bulk graphene Hamiltonian,
$\sigma^{}_{x,y,z,0}$ and $\tau^{}_{x,y,z,0}$ are Pauli
matrices in the sublattice and valley spaces respectively, 
$\vec{p}=-i\hbar\vec{\nabla} + e\vec{A}$
and $v^{}_F$ is the Fermi velocity. The vector potential is taken as
$\vec{A}=(\Phi/2\pi r)\vec{e}_\varphi$ where $\Phi$ is the total
magnetic flux threading the ring. 
The index $N$ stands for the pair of indices $[m,\tau]$, where we assume
the electrons are in the lowest Landau level, $m$ is the orbital angular
momentum, $\tau=+1$ in the $K$ valley and $\tau=-1$ in the $K'$ valley.
We write the wave functions for $V(r)=0$ using the dimensionless length
$\rho=r/W$ and energy $\varepsilon^{}_N = E^{}_N W / \hbar v^{}_F$ where $W$ is
the width of the ring [Fig.~\ref{fig:Wcomp}(c)] as
\begin{gather}
	\label{eq:BulkWF}
	\psi^{}_{N}(\rho) = e^{i(m-\frac{1}{2})\varphi} b^{}_N
	\begin{bmatrix} f^{}_{N}(|\varepsilon^{}_N|\rho) \\ 
		i \s(\varepsilon^{}_N) e^{i\varphi}
		g^{}_N(|\varepsilon^{}_N| \rho)
	\end{bmatrix}
	\intertext{where}
	f^{}_N = \alpha^{}_N H^{(1)}_{\bar{m}-\frac{1}{2}}(|\varepsilon^{}_N| \rho)
		+ H^{(2)}_{\bar{m}-\frac{1}{2}}(|\varepsilon^{}_N| \rho), \notag \\
	g^{}_N = \alpha^{}_N H^{(1)}_{\bar{m}+\frac{1}{2}}(|\varepsilon^{}_N| \rho)
		+ H^{(2)}_{\bar{m}+\frac{1}{2}}(|\varepsilon^{}_N| \rho), \notag \\
	\alpha^{}_N = - \frac{ H^{(2)}_{\bar{m}-\frac{1}{2}}(|\varepsilon^{}_N| \rho^{}_-) 
		+ \tau \s(\varepsilon^{}_N) H^{(2)}_{\bar{m}+\frac12}(|\varepsilon^{}_N|
		\rho^{}_-) }
		{H^{(1)}_{\bar{m}-\frac12}(|\varepsilon^{}_N| \rho^{}_-) +
		\tau\s(\varepsilon^{}_N) H^{(1)}_{\bar{m}+\frac12}
		(|\varepsilon^{}_N| \rho^{}_-) } \notag
\end{gather}
with $\rho^{}_{\pm} = \frac{R}{W}\pm\frac{1}{2}$, $\bar{m} = m +
\frac{\Phi}{\Phi^{}_0}$, $b^{}_N$ is the normalization factor and $\s(x)=1$
for $x\geq0$ and $\s(x)=-1$ for $x<0$. The functions $H^{(1)}_\nu(x)$
and $H^{(2)}_\nu(x)$ are respectively Hankel functions of the first and
second kind. 

The ring confinement is defined by the mass term [written as the
potential $V(r)$] in the Hamiltonian. We employ infinite mass boundary
conditions \cite{bib:Recher-prb76,bib:InfMassBC} so that $V(r)\to
\infty$ outside the ring. This yields the boundary condition
$\psi(\rho^{}_{\pm}) = \tau(\vec{n} \cdot \vec{\sigma})
\psi(\rho^{}_{\pm})$.
The coefficient $\alpha^{}_N$ is found by applying this 
condition at the inside edge, and $b^{}_N$ is calculated numerically for
each state and value of the flux via the normalization condition
for the wave function. 

The interacting few-electron system is studied by adding the term
\begin{equation}
	\label{eq:Coulomb-H}
	C = V(\vec{r}^{}_1 - \vec{r}^{}_2) 
	\tau^{}_0 \otimes \sigma^{}_0 
	= \frac{e^2}{W\epsilon^{}_g} \frac1{\left|\vec{\rho}^{}_1 -
	\vec{\rho}^{}_2\right|} \tau^{}_0\otimes\sigma^{}_0
\end{equation}
to the Hamiltonian where $\epsilon^{}_g$ is the dielectric constant of
monolayer graphene. The simple matrix structure of this operator is
unchanged by the transformation to the valley-symmetric form. 
We evaluate (numerically) the matrix elements of this operator over the
single-particle wave functions in Eq.~\eqref{eq:BulkWF}. 
Using these single-particle matrix elements, we construct the many body
Hamiltonian and carry out an exact diagonalization procedure to
determine the energy and eigenstates of the interacting system. The
persistent current $j$ is then calculated by taking the derivative of
the ground state energy $E^{}_0$ of the few electron system with respect to
the flux as $j(\Phi) = \frac{\partial}{\partial\Phi} E^{}_0$.

\begin{figure*}[tb]
	\centering
	\includegraphics[]{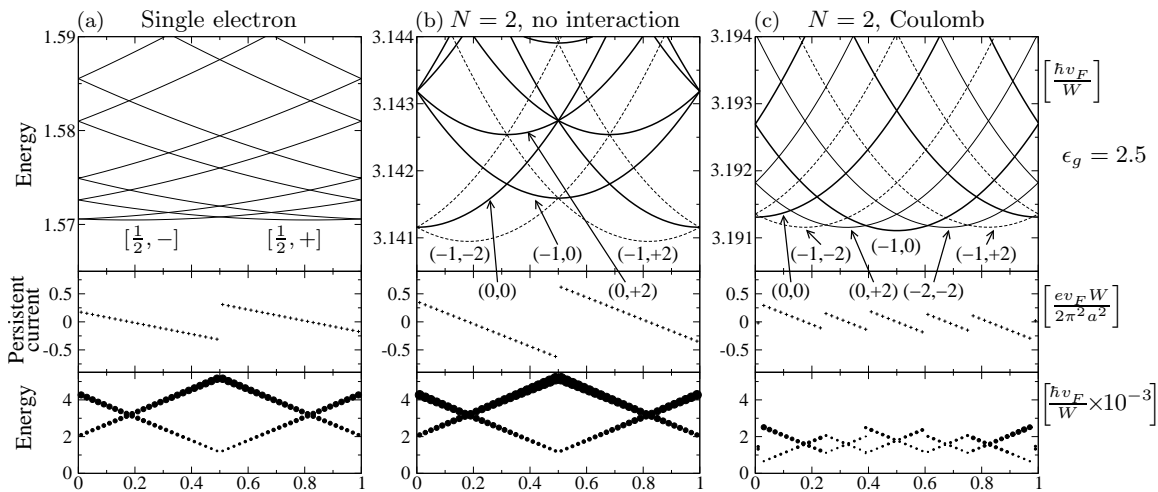}
	\caption{Energy spectrum, persistent current (middle pane) and
	optical absorption of unpolarized light (lower pane) by (a) a single
	electron, (b) two non-interacting electrons, and (c) two electrons
	with the Coulomb interaction included. States
	in the two-electron plots are labelled by the pair of quantum
	numbers $(M,T)$ where $M$ is the total angular momentum, and
	$T$ is the total valley index. The area of
	the points in the absorption plots represent the intensity of the
	peak in arbitrary units. In all three plots, $W=10\,\unit{nm}$ and
	$\frac{R}{W}=10$, and $\epsilon^{}_g=2.5$ \cite{bib:Ando-JPSJ75}.
	\label{fig:N1N2plots}}
\end{figure*}

To describe the absorption of incident light by the graphene ring we add
a term to the Hamiltonian which describes the coupling of electrons to
the field via the vector potential $\vec{A}^{}_{\mathrm{EM}}=2A^{}_0
\vec{\alpha}\cos(\vec{k}\cdot\vec{r}-\omega
t)$. We assume that the radiation propagates as a plane wave with wave
vector $\vec{k}$, frequency $\omega$, and polarization described by the
unit vector $\vec{\alpha}$. Then, the Hamiltonian can be written
\begin{equation}
	\label{eq:H-EM}
	{\cal H} = v^{}_F \vec{\sigma} \cdot \left( \vec{p} + e\vec{A}^{}_{\mathrm{B}} 
		+ e\vec{A}^{}_{\mathrm{EM}} \right) + \tau V(r) \sigma^{}_z + C
\end{equation}
in the valley symmetric representation. The transition rate from state $N$ 
to state $N'$ is calculated from
\begin{equation}
	\label{eq:TR}
	w^{}_{N'N} \propto \left| \left\langle N' \right| 
		\sigma^{}_x \alpha^{}_x + \sigma^{}_y \alpha^{}_y 
		\left| N \right\rangle \right|^2 
	= 4\pi^2 \left| I^{}_{N'N} \right|^2,
\end{equation}
with
\begin{multline}
        \label{eq:abs-I}
        I^{}_{N'N} = \int_{\rho^{}_-}^{\rho^{}_+} \rho \, d\rho \,
                b_{N'}^\ast b^{}_N \left( \delta^{}_{\tau',K} \delta^{}_{\tau,K}
                + \delta^{}_{\tau',K'} \delta^{}_{\tau,K'} \right) \times \\
                \times \Big[ \delta^{}_{m',m+1} (\alpha^{}_x-i\alpha^{}_y) f_{N'}^\ast g^{}_N
                - \delta^{}_{m',m-1} (\alpha^{}_x+i\alpha^{}_y) g_{N'}^\ast f^{}_N \Big]
\end{multline}
in the dipole approximation.
The integral (where we drop the coordinate dependence of the
spatial functions for brevity) must be evaluated numerically.
The intensity of the absorption is proportional to this transition rate 
and the area of the dots in the lowest panels of
Fig.~\ref{fig:N1N2plots}(a), Fig.~\ref{fig:N1N2plots}(b) and
Fig.~\ref{fig:N1N2plots}(c) scale with this quantity. In all figures we
show the absorption of unpolarized light [\textit{i.e.}
$\vec{\alpha}=(\vec{e}_x+\vec{e}_y)/\sqrt{2}$].
Equation~\eqref{eq:abs-I} shows that transitions which change the
angular momentum quantum number by $\pm1$ are permitted, so long as the
valley index remains the same. Where the initial or final states of the
transition are degenerate, we take the average of the intensity of all
possible pairs of initial and final states.


In Fig.~\ref{fig:N1N2plots}(a) we show the energy spectrum, persistent
current and optical absorption for a single electron in the graphene
ring with $R/W=10$. The lifting of the valley degeneracy previously 
described causes the step in the persistent current at 
$\phi=\Phi/\Phi^{}_0=0.5$. 
For $0 < \phi < 0.5$ the ground state consists of one electron in the
$m=-\frac12$, $\tau=-1$ state whereas for $0.5<\phi<1$ the
valley index is $\tau=+1$. 
For $\phi \gtrsim 0$, transitions to the lowest-lying states
$m=+\frac12$, $\tau=+1$ and $m=-\frac12$, $\tau=-1$ are not
allowed since the optical absorption cannot mix valleys. 

For two non-interacting electrons, the ground state consists of a pair
of electrons with anti-parallel spins occupying the same single-particle
states as in the single-electron system [Fig.  \ref{fig:N1N2plots}(b)].
The persistent current reflects the similarity between the ground states
of the single-particle and $N=2$ non-interacting system, and since there
are now two electrons, the persistent current is doubled.
The excited states can have varying degrees of degeneracy: If the
quantum number pairs $P=[m^{}_P,\tau^{}_P]$ and $Q=[m^{}_Q,\tau^{}_Q]$
of the two electrons are identical then there is only one permitted
configuration of the electron spins, the singlet state. However, if
$P\neq Q$ then there are four degenerate possibilities: the singlet and
three triplet states. 

When the Coulomb interaction is included [Fig.~\ref{fig:N1N2plots}(c)], the
picture changes drastically. To describe the two particle states, we
introduce the notation $M=m^{}_1+m^{}_2$ for the total angular momentum and
$T=\tau^{}_1+\tau^{}_2$ for the total valley quantum number. 
The exchange interaction will split the degenerate
singlet-triplet states when both of the electrons are in the same valley
{\it i.e.} for $T=\pm2$. In this case, the energy of the singlet
does not contain any contribution from exchange and consequently
has a rather higher energy than the corresponding triplet. This is
exemplified by the $(M=0, T=2)$ state. The triplet part
experiences exchange and this reduces its energy sufficiently for it to
form the ground state for $\phi\approx0.3$ with
$\varepsilon\simeq3.191$. At the same flux the singlet state has
$\varepsilon\simeq3.205$ and is therefore not present in
Fig.~\ref{fig:N1N2plots}(c). 
On the other hand, the singlet and triplet parts of the $(-1,0)$ 
degenerate state are not split by the exchange interaction.

Because the Coulomb matrix elements depend on the angular momenta of the
single-electron states involved, the size
of the energy change will vary between different states. This is shown
in the vicinity of the crossing of the lowest three states for
$\phi=0$.
The $M=\pm1$ states each contain two electrons in the same angular
momentum state ($m=\pm\frac{1}{2}$) so their interaction is stronger
than the electrons in the $M=0$ state which has electrons in different
angular momentum states. This causes the ground state to become a
degenerate singlet-triplet combination for a small range of flux.

This intricate interplay of different-sized contributions from the
Coulomb interaction adds significant complexity to the ground state of
the interacting system. In the non-interacting case the ground state is
always comprised of a singlet, but the 
interaction introduces several additional level crossings which give
rise to ranges of the flux where the ground state becomes a triplet or
degenerate singlet-triplet state. Moreover, since the energy due to the
Coulomb interaction depends on the angular momentum of the state, the
size of the ring will also be an important factor. In Fig.~\ref{fig:Wcomp}
(a) we plot the ground state energy for $\frac{R}{W}=3$
(a wide ring) and $\frac{R}{W}=25$ (a narrow ring) to illustrate this
dependence. 
The relative depth of the minima of the ground state energy vary, and
the nature of the ground state at zero field changes from the $(-1,-2)$
singlet in a wide ring to the $(0,0)$ singlet-triplet in a narrow ring.
This transition is revealed by the absorption spectrum since the
crossover to the degenerate ground state changes the spectrum to two
closely-spaced low intensity peaks. 
In Fig. \ref{fig:Wcomp}(b), the difference in energy between the $(0,0)$
and $(-1,-2)$ is plotted as a function of $\frac{R}{W}$ for a ring with
$R=100\,\unit{nm}$. The crossover for the ground state occurs at
approximately $\frac{R}{W}=7$, independent of the value of $R$.

\begin{figure}[tb]
	\centering
	\includegraphics{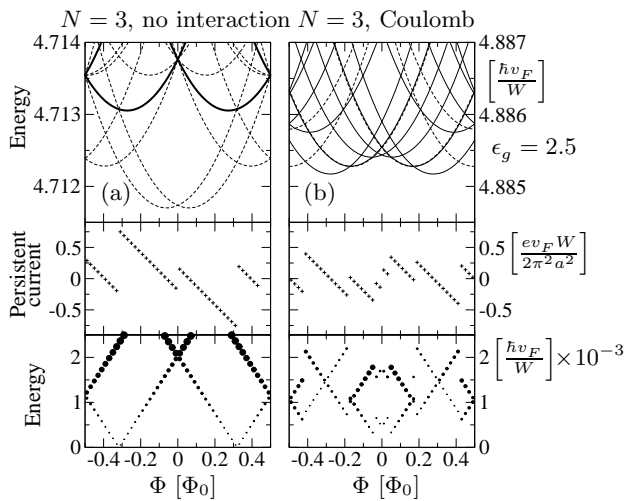}
	\caption{(a) Non-interacting, and (b) interacting three electron
        energy spectrum, persistent current and optical absorption for
	$\frac{R}{W}=10$. Dashed lines denote two-fold degeneracy, solid
	lines four-fold degeneracy and thick solid line eight-fold
	degeneracy of the state.
	\label{fig:N3-int}}
\end{figure}

For three non-interacting electrons in the ring, the ground
state is composed of spin and valley unpolarized states ({\it i.e.}
$T=\pm1$).  When the interaction is added, the contribution from
exchange is largest for $T=\pm3$ states so the low energy spectrum
becomes much more compact, just as in the $N=2$ case. Qualitatively, the
effect of the interaction is the same as previously, so that the
changing nature of the ground state again demonstrates the complexity
due to the absence of the valley degeneracy. However, because there are
more possible combinations of states, the persistent current and
absorption spectrum are correspondingly more complex in their structure.
In particular it is not possible to have $T=0$ so the exchange energy is
always finite. However, its contribution is larger for $T=\pm3$ states
than for $T=\pm1$ states.
It is also the case that the width of the ring (and hence the relative
strength of the interaction) will affect the detail of the ground state.

To summarize, we have studied the effect of the electron-electron
interaction on measurable quantities in a graphene quantum ring.  We
find that the interplay of the interaction and the total valley quantum
number allow for an intricate manifestation of the breaking of valley
degeneracy in this geometry. The change of the interacting ground state
between singlet, triplet and degenerate singlet-triplet natures reveals
the sensitivity of the exchange contribution to the total valley index.
These changes in the ground state are manifested in the fractional
nature of the AB oscillations in the persistent current, and in the
steps and intensity changes in the absorption spectrum as the flux is
varied.

This work has been supported by the Canada Research Chairs Program and
the NSERC Discovery Grant, and we would like to thank P. Pietil\"ainen
for helpful discussions.

\end{document}